\documentclass[12pt]{iopart}
\usepackage{graphicx}
\eqnobysec
\renewcommand{\theequation}{\thesection.\arabic{equation}}

\begin{document}

\title[Multi-lump solutions of KP equation with integrable boundary]{Multi-lump solutions of KP equation with integrable boundary via  $\overline\partial$-dressing method}
\author{V. G. Dubrovsky, A. V. Topovsky}
\address{Novosibirsk State Technical University, Karl Marx prospect 20, 630072, Novosibirsk, Russia.}
\ead{\mailto{dubrovsky@ngs.ru},\mailto{dubrovskij@corp.nstu.ru}, \mailto{topovskij@corp.nstu.ru}, \mailto{topovsky.av@gmail.com}}



\begin{abstract}
We constructed the new classes of exact multi-lump solutions of KP-1 and KP-2 versions of KP equations with integrable boundary condition $u_{y}\big|_{y=0}=0$ by the use of $\overline\partial$-dressing method of Zakharov and Manakov and derived general determinant formula for such solutions. We demonstrated how reality and boundary conditions for the field $u$ in the framework of $\overline\partial$-dressing method can be exactly satisfied. Here we present explicit examples of two-lump solutions with integrable boundary as nonlinear superpositions of two more simpler "deformed"
one-lump  solutions:
 the fulfillment of boundary condition leads to formation of certain
 eigenmodes of the field $u(x,y,t)$ in semiplane $y\geq 0$ as analogs of standing waves on string with fixed end points.
\end{abstract}

\noindent{\it Keywords\/}: KP equation, $\overline{\partial}$-dressing method, lump solutions, integrable boundaries
\pacs{02.30.Ik, 02.30.Jr, 02.30.Zz, 05.45.Yv}

\section{Introduction}
\label{Section_1}
The famous KP-equation~\cite{KadPetv}:
\begin{equation}\label{KP}
u_{t}+ u_{xxx}+6uu_x+3\sigma^2\partial_x^{-1}u_{yy}=0,
\end{equation}
with $\sigma=i$ for KP-1 and $\sigma=1$ for KP-2, can be represented as compatibility condition in the well known Lax form $\left[L_1,L_2\right]=0$ of two linear auxiliary problems~\cite{Dryuma}-\cite{ZakharovShabat2}:
\begin{equation}\label{KP auxiltary problems}
\left\{
\begin{array}{ll}
L_1\psi=(\sigma\partial_y+\partial^2_x+u)\psi=0, \\
L_2\psi=(\partial_t+4\partial^3_x+6u\partial_x+3u_x-3\sigma^2\partial^{-1}_xu_y)\psi=0.
\end{array}
\right.
\end{equation}
The first linear problem (\ref{KP auxiltary problems}) with $y$ as \,"time"\, variable represents nonstationary Schroedinger equation - for KP-1 case with $\sigma=i$ and diffusion or heat equation - for KP-2 case with $\sigma=1$, respectively.

KP-equation (\ref{KP}) as the first remarkable integrable example
from known now long list of 2+1-dimensional integrable nonlinear
equations is well studied: several classes of exact solutions,
hamiltonian and recursion structures, Cauchy problem, etc. for this equation have been extensively investigated, see, for example,  ~\cite{Manakov}-\cite{Beals&Coifman2} and  ~\cite{KonopelchenkoBook1}, ~\cite{KonopelchenkoBook2}.

In the present paper we constructed new classes of exact
multi-lump solutions of KP equation (\ref{KP}) with integrable
boundary condition~\cite{Sklyanin}-\cite{Habibullin2}
\begin{equation}\label{BoundaryCondition}
u_y(x,y,t)=\frac{\partial u(x,y,t)}{\partial y}\bigg|_{y=0}=0
\end{equation}
for both versions KP-1 and KP-2 via $\overline\partial$-dressing
method of Zakharov and Manakov~\cite{Manakov}.
The concept of integrable boundary conditions compatible with
integrability for integrable nonlinear equations was first introduced by Sklyanin~\cite{Sklyanin}.
In subsequent papers of Habibullin et al
~\cite{Habibullin}-\cite{HabibullinKDV} and others ~\cite{Vereshchagin} this concept of integrable boundary conditions to several types of integrable nonlinear equations has been applied: for difference equations, 1+1-dimensional and 2+1-dimensional integrable nonlinear differential and integro-differential  equations. In these papers a list of integrable boundary conditions for known 2+1-dimensional nonlinear equation such as KP, mKP, Nizhnik-Veselov-Novikov, Ishimori and so on with some examples of corresponding solutions have been proposed and calculated ~\cite{Habibullin}-\cite{Vereshchagin}.
A. S. Fokas et al obtained interesting results via so called Unified Approach to Boundary Value problems, see book \cite{FokasBook}, where they demonstrated the applicability of this method for one-dimensional and multi-dimensional linear and nonlinear differential equations.

We demonstrated here how the boundary condition
(\ref{BoundaryCondition}) and condition of reality $u=\overline u$
in construction of exact multi-lump solutions of KP equation
(\ref{KP}) can be in the framework of $\overline\partial$-dressing
method exactly satisfied. We derived the restrictions from reality and integrable boundary conditions to the kernel of  $\overline\partial$-problem in explicit form by the use of general determinant formulas for exact solutions.

The paper is organized as follows. The first section is introduction.  In second section  we are reviewed the basic formulae of $\overline\partial$-dressing for KP equation (\ref{KP}), derived general determinant formula in convenient form for calculations of multi-lump solutions for KP equation and the restrictions on kernel from boundary condition in the "limit of weak fields". In the following third and fourth sections we constructed new classes of multi-lump solutions for KP-1 and KP-2 versions of KP equation. Furthermore, we illustrated the new classes of exact solutions by the simple examples of two-lump solutions. We showed that integrable boundary condition $u_y\big|_{y=0}=0$ leads to formation of bound state of several single lumps as certain eigenmodes for the field $u(x,y,t)$ in semiplane $y\geq0$,  the analogs of standing waves on the string arising from corresponding boundary conditions at endpoints of string.

\section{Basic formulas of $\overline\partial$-dressing for KP equation. Determinant formula for multi-lump solutions. Restrictions from integrable boundary condition}
\label{Section_2}

The general formulas of $\overline\partial$-dressing method of Zakharov and Manakov in application for KP equation are extensively described in~\cite{Zakharov}-\cite{KonopelchenkoBook2}. Central object of $\overline\partial$-dressing is a wave function $\chi(\lambda,\overline\lambda; x,y,t)$ which is the function of spectral variables $\lambda$, $\overline\lambda$ and spacetime variables $x,y,t$. This function is connected with wave function $\psi(\lambda,\overline\lambda; x,y,t)$ of linear auxiliary problems
(\ref{KP auxiltary problems}) for KP (\ref{KP}) by the formula:
\begin{equation}\label{Psi&Chi}
  \Psi(\lambda,\overline\lambda; x,y,t):=\chi(\lambda,\overline\lambda; x,y,t)\exp{F(\lambda;x,y,t)}
\end{equation}
with phase $F(\lambda;x,y,t)$ in exponent:
\begin{equation}\label{F(lambda)}
  F(\lambda; x,y,t)=i\lambda x+\frac{\lambda^2}{\sigma}y+4i\lambda^3t.
\end{equation}
Herein, instead of full notations $\chi(\lambda,\overline\lambda; x,y,t)$, $F(\lambda;x,y,t)$ , etc. we are using short notations with restriction to corresponding dependence on spectral variables $\lambda$, $\overline\lambda$ (or more shorter - on $\lambda$): $\chi(\lambda,\overline\lambda)$ or $\chi(\lambda)$; $F(\lambda)$; $\psi(\lambda,\overline\lambda)$ or $\psi(\lambda)$, etc.

Basic equation of $\overline\partial$-dressing method is the so called $\overline\partial$-problem for wave function $\chi(\lambda,\overline\lambda)$~\cite{KonopelchenkoBook1},~\cite{KonopelchenkoBook2} or equivalently singular integral equation for $\chi$:
\begin{equation}\label{di_problem1}
\fl\chi (\lambda,\overline\lambda) = 1 + \frac{2i}{\pi}\int\int\limits_C {\frac{d{\lambda }'_R
d{\lambda'}_I}{(\lambda'-\lambda)}}
\int\int\limits_C  {d\mu_R  d\mu_I} \chi(\mu,\overline{\mu})
R_0(\mu ,\overline \mu ;\lambda',\overline {\lambda' })e^{F(\mu)-F(\lambda')}
\end{equation}
with kernel $R_0(\mu,\overline\mu;\lambda,\overline\lambda)$ (in short $R_0(\mu,\lambda)$); here as usual $\lambda=\lambda_R+i\lambda_I$, $\mu=\mu_R+i\mu_I$ are the notations for complex spectral variables. In (\ref{di_problem1}) due to (\ref{KP auxiltary problems}), (\ref{Psi&Chi}), (\ref{F(lambda)}) the canonical normalization $\chi\big|_{\lambda\rightarrow\infty}\rightarrow1$ for wave function $\chi (\lambda,\overline\lambda)$ is chosen and will be used herein.  Reconstruction formula for solutions $u(x,y,t)$ of KP equation (\ref{KP})
\begin{equation}\label{reconstruct}
u(x,y,t)=-2i\partial_x \chi_{-1}(x,y,t)
\end{equation}
expresses $u$ through the $\chi_{-1}$ coefficient of Taylor expansion of $\chi(\lambda,\overline\lambda)$ in the neighborhood of $\lambda=\infty$:
\begin{equation}\label{TaylorExp}
\chi(\lambda,\overline\lambda)=1+\frac{\chi_{-1}}{\lambda}+\ldots .
\end{equation}
Formula (\ref{reconstruct}) is valid for both versions: $\sigma=i$, KP-1 and $\sigma=1$, KP-2 of (\ref{KP}). From (\ref{di_problem1}) one has for $\chi_{-1}$:
\begin{equation}\label{chi_(-1)}
\chi_{-1}(x,y,t) = -\frac{2i}{\pi}\int\int\limits_C d\lambda_R d\lambda_I
\int\int\limits_C  \chi(\mu,\overline{\mu})R_0(\mu,\lambda)e^{F(\mu)-F(\lambda)}d\mu_R  d\mu_I.
\end{equation}

For delta-form kernel $R_0$ of the type (the sum of products of delta-functions)
\begin{equation}\label{kernel1}
R_0(\mu,\overline{\mu};\lambda,\overline{\lambda})
=\sum\limits_{k=1}^{N} A_k\delta(\mu-\mu_k)\delta(\lambda-\mu_k)
\end{equation}
with complex amplitudes $A_k$ and complex \,"spectral"\, points $\mu_k$ one can easily obtain general determinant formula for exact multi-lump  solutions (complex in general) for KP equation (\ref{KP}). We repeated the derivation of well known determinant formula for multi-lump solutions~\cite{KonopelchenkoBook1},~\cite{KonopelchenkoBook2}  introducing by the way convenient notations and useful terminology. From (\ref{di_problem1}) and (\ref{kernel1}) one obtains the wave function $\chi(\lambda,\overline\lambda)$
\begin{equation}\label{chi(lambda)}
\chi(\lambda,\overline\lambda)=1-\frac{2i}{\pi}\sum^N_{k=1}
\frac{A_k}{\lambda-\mu_k}\chi(\mu_k)
\end{equation}
in the form of the sum of $N$ terms with simple poles at \,"spectral"\, points $\mu_k$; such pole structure of wave function on spectral variables $\lambda$ is typical for quantum mechanics with basic Schr\"{o}dinger equation and corresponding pole structures of quantum-mechanical wave functions from wave numbers, energy, momentum, etc. Formula (\ref{chi(lambda)}) expresses the wave function  $\chi(\lambda,\overline\lambda)$ at arbitrary values of spectral variables $\lambda, \overline\lambda$ in terms of some kind of "basis" or basic set of $N$ wave functions $\chi(\mu_k):=\chi(\mu_k,\overline\mu)$, $(k=1,\ldots,N)$ at spectral points $\mu_k$ corresponding to the choice (\ref{kernel1}) of the kernel $R_0$.

From (\ref{F(lambda)}), (\ref{di_problem1}) and (\ref{chi(lambda)}) one obtains linear algebraic system for the wave functions $\chi(\mu_k)$:
\begin{equation}\label{tildeA}
\sum\limits_{l=1}^N\tilde{A}_{kl}\chi(\mu_l)=1;\quad \tilde{A}_{kl}:=\left(1-\frac{2A_k}{\pi}\tilde{\Phi}_k\right)\delta_{kl}-
\frac{2iA_l}{\pi(\mu_l-\mu_k)}(1-\delta_{kl}),
\end{equation}
here we introduced the quantities $\tilde{\Phi}_k$ due to (\ref{F(lambda)}) by definitions:
\begin{equation}\label{dF_formula}
F'(\mu_k):=\frac{\partial F(\mu)}{\partial\mu}\bigg|_{\mu=\mu_k}=i\left(x-\frac{2i\mu_k}{\sigma}y+12\mu^2_k t\right):=i\tilde{\Phi}_k(\mu_k).
\end{equation}
 Factoring from the matrix $\tilde{A}_{kl}$ diagonal matrix $D_{kl}:=-\frac{2A_l}{\pi}\delta_{kl}$ we introduced instead of $\tilde{A}_{kl}$ more simple and convenient matrix $A_{kl}$:
\begin{equation}\label{AD}
A_{kl}:=(\tilde{A}D^{-1})_{kl}=\Phi_{k}\delta_{kl}-
\frac{i(1-\delta_{kl})}{\mu_k-\mu_l},
\end{equation}
where
\begin{equation}\label{XtildeX}
\Phi_{k}(\mu_k):=\tilde \Phi_k(\mu_k)-\frac{\pi}{2A_k}=x-\frac{2i\mu_k}{\sigma}y+12\mu^2_k t-\gamma_k,\quad \gamma_k:=\frac{\pi}{2A_k}.
\end{equation}
So all dependence on amplitudes $A_k$  is transferred to constants $\gamma_k=\frac{\pi}{2A_k}$ in phases $\Phi_{k}$.

Using (\ref{reconstruct}), (\ref{chi_(-1)})-(\ref{AD}) we derived the exact multi-lump solution of KP equation
\begin{equation}
u(x,y,t)=-2i\partial_x\chi_{-1}=2\frac{\partial}{\partial x}\sum\limits_{k,l}A^{-1}_{kl}
\end{equation}
with matrix $A_{kl}$ given by (\ref{AD}). Due to special structure of matrix $A_{kl}$ we proved that
\begin{equation}
\sum\limits_{k\neq l}A^{-1}_{kl}=0,
\end{equation}
and hence we obtained
\begin{eqnarray}\label{ReconstructFinal}
 u(x,y,t)=2\frac{\partial}{\partial x}\sum\limits_{k,l=1}^{N}A^{-1}_{kl}=2\frac{\partial}{\partial x}\sum\limits_{k=1}^{N}A^{-1}_{kk}=
2\frac{\partial}{\partial x}\sum\limits_{k,l=1}^{N}A_{lk,x}A^{-1}_{kl}=\nonumber\\
=2\frac{\partial}{\partial x}\textrm{tr}(A_xA^{-1})=2\frac{\partial^2}{\partial x^2}\ln\det A.
\end{eqnarray}
Here we used in derivation (\ref{ReconstructFinal}) famous relation $\frac{\partial}{\partial x}\ln\det A=\textrm{tr}(A_xA^{-1})$. Formula (\ref{ReconstructFinal}) gives in general determinant form exact  multi-lump solutions (complex in general) of KP equation for both versions KP-1 and KP-2.

In order to satisfy reality $\overline u=u$ and integrable boundary  (\ref{BoundaryCondition}) conditions we had to specifically choose complex constant amplitudes $A_k$ and complex spectral points $\mu_k$. In fact, this is the main problem in calculations of exact real solutions with integrable boundary condition (\ref{BoundaryCondition}).
The integrable boundary condition (\ref{BoundaryCondition}) can be satisfied in the framework of $\overline\partial$-dressing method as follows. From reconstruction formula (\ref{reconstruct}) by the use of expression (\ref{chi_(-1)}) for the coefficient $\chi_{-1}$ of expansion (\ref{TaylorExp}) we derived taking into account (\ref{F(lambda)}) and (\ref{di_problem1}):
\begin{eqnarray}\label{BoundaryConditionWeakField1}
\fl u_y\big|_{y=0}=-2i\frac{\partial^2\chi_{-1}}{\partial x\partial y}\big|_{y=0}\cong\nonumber\\
\fl\frac{4}{\pi}\int\int\limits_{C}d\lambda_{R}d\lambda_{I}
\int\int\limits_{C}(\mu-\lambda)(\mu^2-\lambda^2)
R_0(\mu,\overline\mu;\lambda,\overline\lambda)e^{F(\mu)-
F(\lambda)}\big|_{y=0}d\mu_{R}d\mu_{I}.
\end{eqnarray}
Here, in the "limit of weak fields", we chose for the wave function $\chi$ in integrand approximate value $\chi(\mu,\overline\mu)\approx1$ as first iteration from (\ref{di_problem1}). Under the change of variables $\mu\leftrightarrow-\lambda$:
\begin{equation}
(F(-\lambda)-F(-\mu))\big|_{y=0}=(F(\mu)-F(\lambda))\big|_{y=0},
\end{equation}
and the expression (\ref{BoundaryConditionWeakField1}) transforms to the following one
\begin{equation}\label{BoundaryConditionWeakField2}
\fl u_y\big|_{y=0}\cong-\frac{4}{\pi}\int\int\limits_{C}d\mu_{R}d\mu_{I}\int\int\limits_{C}(\mu-\lambda)(\mu^2-\lambda^2)
R_0(-\lambda,-\overline\lambda;-\mu,-\overline\mu)e^{F(\mu)-F(\lambda)}\big|_{y=0}d\lambda_{R}d\lambda_{I}.
\end{equation}
Imposing on $R_0(\mu,\overline\mu;\lambda,\overline\lambda)$ the restriction
\begin{equation}\label{BoundaryConditionWeakField}
R_0(\mu,\overline\mu;\lambda,\overline\lambda)=
R_0(-\lambda,-\overline\lambda;-\mu,-\overline\mu),
\end{equation}
we derived from (\ref{BoundaryConditionWeakField1})-(\ref{BoundaryConditionWeakField}) that
\begin{equation}
u_y\big|_{y=0}=-u_y\big|_{y=0}=0
\end{equation}
and consequently satisfied the  boundary condition (\ref{BoundaryCondition}).
The restriction (\ref{BoundaryConditionWeakField}) on the kernel $R_0$ is obtained in the "limit of weak fields" (we chose for the wave function $\chi(\mu,\overline\mu)$ under integrand (\ref{BoundaryConditionWeakField1}) the first iteration $\chi(\mu,\overline\mu)\cong1$).

In spite of non rigorous character of  derivation (\ref{BoundaryConditionWeakField}) this restriction can be successfully applied for choosing appropriate kernels $R_0(\mu,\overline\mu;\lambda,\overline\lambda)$ in calculations of real exact solutions of KP equations (\ref{KP}) with integrable boundary condition (\ref{BoundaryCondition}). We performed these calculations in three following steps:

\begin{enumerate}
\item the derivation of restrictions on kernel $R_0$  from reality  condition $u=\overline {u}$  for complex-valued solutions via general determinant form;
\item appropriate choice of the kernel $R_0$ (amplitudes $A_k$, spectral points $\mu_k$) in accordance with restriction from reality and restriction (\ref{BoundaryConditionWeakField}) from boundary condition;
\item the calculation by reconstruction formula (\ref{ReconstructFinal}) of exact real multi-lump solutions of KP equation.
  \end{enumerate}

\section{Multi-lump solutions of KP-1 equation with integrable boundary condition}
\label{Section_3}
At first we calculated multi-lump solution of KP-1 equation with integrable boundary (\ref{BoundaryCondition}) corresponding to delta-form kernel (\ref{kernel1}) with real spectral points $\mu_k=\overline\mu_k$. To restriction (\ref{BoundaryConditionWeakField}) from integrable boundary condition (\ref{BoundaryCondition}) satisfies delta-form (\ref{kernel1}) kernel with paired terms:
\begin{equation}\label{Kernel2}
R_0(\mu,\overline\mu;\lambda,\overline\lambda)=\sum\limits_{k=1}^N\left(a_k\delta(\mu-\mu_k)\delta(\lambda-\mu_k)+
a_k\delta(\mu+\mu_k)\delta(\lambda+\mu_k)\right).
\end{equation}
To apply general determinant formula (\ref{ReconstructFinal})  with kernel $R_0$ (\ref{Kernel2})  it is convenient to rewrite this kernel in form (\ref{kernel1})
\begin{equation}\label{KernelGenForm}
R_0(\mu,\overline{\mu};\lambda,\overline{\lambda})
=\sum\limits_{k=1}^{2N} A_k\delta(\mu-M_k)\delta(\lambda-M_k),
\end{equation}
with sets of amplitudes $A_{k}$ and spectral points $M_{k}$, $k=1,\ldots,2N$:
 \begin{eqnarray}\label{setA&M_KP1}
(A_1, A_2,\ldots,A_{2N-1},A_{2N}):=(a_1,a_1;a_2, a_2;\ldots,a_N,a_N),\nonumber \\
(M_1,M_2,\ldots,M_{2N-1},M_{2N}):=(\mu_1,-\mu_1;\mu_2,-\mu_2,\ldots,\mu_N,-\mu_N).
\end{eqnarray}
Phases $\Phi_k(\pm\mu_k)$ due to formulas (\ref{XtildeX}) have the form:
 \begin{eqnarray}\label{XtildeX1}
\Phi_k(\pm\mu_k)=x+12\mu_k^2t\mp2\mu_k y-\gamma_k:=X_k(x,t)\mp Y_k(y),\nonumber \\
X_k(x,t):=x+12\mu^2_k t-\gamma_k,\quad Y_k(y)=\overline{Y_k(y)}:=2\mu_k y,\quad \gamma_k=\frac{\pi}{2a_k}.
\end{eqnarray}
Matrix $A$ due to (\ref{AD}) and (\ref{setA&M_KP1}) has the following diagonal $A_D$ and non diagonal $A_F:=A-A_D$ parts:
\begin{equation}\label{ADAND}
A=A_D+A_F,\quad (A_F)_{kl}=-\frac{i(1-\delta_{kl})}{M_k-M_l}, \quad (k,l=1,...,2N),
\end{equation}
here the diagonal part of $A_{kl}$ has the form:
\begin{equation}
A_{D}=\textrm{diag}\left(X_1-Y_1,X_1+Y_1;\ldots;X_N-Y_N,X_N+Y_N\right).
\end{equation}
Evidently $A^{+}=A$ ($A^{+}$ denotes hermitian conjugate to $A$) if $A_D=\overline{A_D}$, i.e.
\begin{equation}
\gamma_k=\overline{\gamma_k} \Rightarrow \overline{a}_k=a_k,\quad (k=1,\ldots,N).
\end{equation}
So for real amplitudes $a_k$ in (\ref{KernelGenForm}):
\begin{equation}\label{real_detA}
  \det A=\overline{\det A},
\end{equation}
moreover, according to special structure (\ref{ADAND}) of matrix $A$ with (\ref{XtildeX1})
$\det A$ is even function of $y$ and due to this fact
\begin{equation}\label{EvenFunct}
\det A=\det A(y^2)\Rightarrow(\det A)_y\big|_{y=0}=0.
\end{equation}
Reconstruction formula (\ref{ReconstructFinal}), with the properties (\ref{ADAND})-(\ref{EvenFunct})
of parameters of matrix $A$, gives determinant formula for exact real multi-lump solution of KP-1 equation with integrable boundary (\ref{BoundaryCondition}).

\begin{figure}[h]
\begin{center}
\includegraphics[width=0.50\textwidth, keepaspectratio]{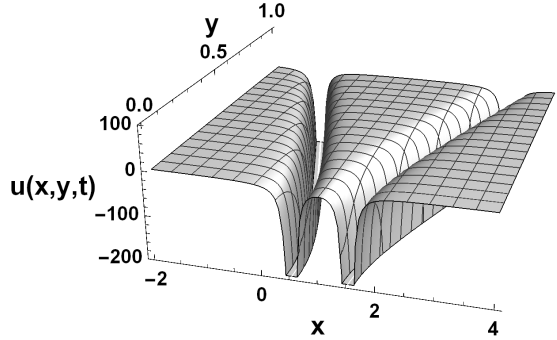}
\parbox[t]{1\textwidth}{\caption{Two-lump solution of KP-1 $u$ (\ref{LumpSolution1})  with parameter $\mu_{1}=1$, $\gamma=1$.}\label{Lump(RealSpectral)KP1}}
\end{center}
\end{figure}

As illustration we calculated the simplest two-lump solution $u(x,y,t)$ of KP-1 equation corresponding to one paired terms $N=1$ in (\ref{Kernel2}). Matrix $A$ (\ref{ADAND}) in this case has the form:
\begin{equation}
 A= \left(
  \begin{array}{cc}
    X_1(x,t)-Y_1(y) & -\frac{i}{2\mu_1} \\
    \frac{i}{2\mu_1} & X_1(x,t)+Y_1(y) \\
  \end{array}
\right),
\end{equation}
and  reconstruction formula (\ref{ReconstructFinal}) leads to the following exact two-lump solution
\begin{eqnarray}\label{LumpSolution1}
u(x,y,t)=2\frac{\partial^2}{\partial x^2}\ln\det A=\nonumber\\
=-\frac{2}{\left(X_1(x,t)-\sqrt{Y_1^2(y)+\frac{1}{4\mu^2_1}}\right)^2}-
\frac{2}{\left(X_1(x,t)+\sqrt{Y_1^2(y)+\frac{1}{4\mu^2_1}}\right)^2}
\end{eqnarray}
with integrable boundary condition (\ref{BoundaryCondition}). The graph of this solution on semi-plane $y\geq 0$ is shown on figure (\ref{Lump(RealSpectral)KP1}).
This solution evidently is singular on the plane $(X_1,Y_1)$ with singularity lines in the form of hyperbolas:
\begin{equation}
X_1^2(x,t)-Y_1^2(y)=\frac{1}{4\mu^2_1}.
\end{equation}

It is interesting to note that to simple kernels $R_{01}$, $R_{02}$ (first and second terms in (\ref{Kernel2})) of the forms
\begin{equation}
R_0(\mu,\lambda)=a_1\delta(\mu-\mu_1)\delta(\lambda-\mu_1),\quad R_0(\mu,\lambda)=a_1\delta(\mu+\mu_1)\delta(\lambda+\mu_1),
\end{equation}
due to general formulas (\ref{tildeA})-(\ref{ReconstructFinal}), correspond singular one-lump solutions:
\begin{equation}\label{OneLumpSolution1}
\fl u_1(x,y,t)\big|_{\mu=\mu_1}=-\frac{2}{\left(X_1(x,t)-Y_1(y)\right)^2},\quad u_2(x,y,t)\big|_{\mu=-\mu_1}=-\frac{2}{\left(X_1(x,t)+Y_1(y)\right)^2}
\end{equation}
with the same $X_1(x,t)$ and $Y_1(y)$ as in (\ref{LumpSolution1}) given by  (\ref{XtildeX1}). The sum of one-lump solutions (\ref{OneLumpSolution1}) is very similar to exact two-lump solution (\ref{LumpSolution1}) with integrable boundary (\ref{BoundaryCondition}). The solutions (\ref{OneLumpSolution1}) (one-lump type) separately  did not satisfy to integrable boundary condition (\ref{BoundaryCondition}), but the exact two-lump solution (\ref{LumpSolution1}) is even function of $y$ and due to this fact  satisfies to (\ref{BoundaryCondition}). Via to (\ref{LumpSolution1}) - (\ref{OneLumpSolution1}) we rewrote (\ref{LumpSolution1}) in the form:
\begin{equation}\label{LumpSolution12}
\fl u(x,y,t)=\left(u_1+u_2\right)\big|_{Y_1\rightarrow Y_{1D}=\sqrt{Y_1^2+\frac{1}{4\mu^2_1}}}=
-\frac{2}{\left(X_1-Y_{1D}\right)^2}-\frac{2}{\left(X_1+Y_{1D}\right)^2}.
\end{equation}
The exact two-lump solution (\ref{LumpSolution1}) or (\ref{LumpSolution12}) with boundary condition (\ref{BoundaryCondition}) can be interpreted as "bound state" of two one-lump solutions (\ref{OneLumpSolution1}) with\, "deformed"\, phase $Y_{1D}=\sqrt{Y^2_1(y)+\frac{1}{4\mu_1^2}}$; the effect of \,"interaction"\, of simple lumps (\ref{OneLumpSolution1}) consists in nonlinear phase shift $Y_{1D}\rightarrow Y_{1D}=\sqrt{Y^2_1(y)+\frac{1}{4\mu_1^2}}$.
We concluded that imposition of boundary condition (\ref{BoundaryCondition}) leads to the formation of \,"eigenmode"\, of the field $u(x,y,t)$ in semiplane $y\geq0$ in the form of two bounded lumps $u_1$ and $u_2$ (\ref{OneLumpSolution1}); this \,"eigenmode"\, of $u(x,y,t)$ with two bounded lumps given by (\ref{LumpSolution1}) reminds standing wave on elastic string arising from corresponding boundary conditions at the ends of the string.

Another interesting class of multi-lump solutions of KP-1 equation corresponds to the kernel similar to (\ref{Kernel2}) but with pure imaginary spectral points $i\mu_{k0}$ and $(-i\mu_{k0})$:
\begin{equation}\label{Kernel3}
\fl R_0(\mu,\overline\mu;\lambda,\overline\lambda)=
\sum\limits_{k=1}^N\left(a_k\delta(\mu-i\mu_{k0})\delta(\lambda-i\mu_{k0})+
a_k\delta(\mu+i\mu_{k0})\delta(\lambda+i\mu_{k0})\right).
\end{equation}
Such kernel satisfies the restriction (\ref{BoundaryConditionWeakField}) from boundary condition (\ref{BoundaryCondition}) and, as it shown further,  for real amplitudes $a_k$
\begin{equation}\label{ConditionOnAmplitude}
\overline{a}_k=a_k,\quad (k=1,\ldots,N)
\end{equation}
reconstruction formula (\ref{ReconstructFinal}) by the use of (\ref{Kernel3}) gives real exact solutions of KP-1 with integrable boundary (\ref{BoundaryCondition}).
In order to apply general determinant formula (\ref{ReconstructFinal}) in considered case it is convenient to rewrite (\ref{Kernel3}) in form (\ref{KernelGenForm}), where sets of amplitudes $A_k$ and spectral points $M_k$ $(k=1,2,\ldots,2N)$ have the form:
 \begin{eqnarray}\label{setA&M_KP1(1)}
\fl (A_1, A_2,\ldots,A_{2N-1},A_{2N}):=(a_1,a_1;a_2, a_2;\ldots,a_N,a_N),\nonumber \\
\fl (M_1,M_2,\ldots,M_{2N-1},M_{2N}):=
(i\mu_{10},-i\mu_{10};i\mu_{20},-i\mu_{20},\ldots,i\mu_{N0},-i\mu_{N0}).
\end{eqnarray}
Phases $\Phi_k$ for real amplitudes $a_k$, due to formulas (\ref{XtildeX}) and (\ref{ConditionOnAmplitude}), (\ref{setA&M_KP1(1)}), have the form:
\begin{eqnarray}\label{X_kp1}
\Phi_k(\pm i\mu_{k0})=x-12\mu^2_{k0}t-\gamma_k\mp i2\mu_{k0}y:=X_k(x,t)\mp iY_k(y), \nonumber\\
\gamma_k=\overline{\gamma}_k=\frac{\pi}{2a_k},\quad
\overline{X}_k(x,t)=X_k(x,t),\quad \overline{Y}_k(y)=Y_k(y).
\end{eqnarray}

Due to (\ref{ConditionOnAmplitude})-(\ref{X_kp1})  matrix $A$ has the following diagonal $A_D$ and non diagonal $A_F:=A-A_D$ parts:
\begin{eqnarray}\label{ADAND1}
A=A_D+A_{F},\quad (A_{F})_{kl}=-\frac{i(1-\delta_{kl})}{M_k-M_l}, \nonumber
\\
A_{D}=\textrm{diag}\left(X_1-iY_1,X_1+iY_1;
\ldots,X_N-iY_N,X_N+iY_N\right).
\end{eqnarray}
From this form of matrix $A$, under requirement of real amplitudes $a_k=\overline {a_k}$ in (\ref{Kernel3}), follow its important properties:
 \begin{equation}\label{detA}
\overline{\det{A}}=\det{A},\quad \det{A}=\det{A}(y^2),
\end{equation}
and consequently the reconstruction formula (\ref{ReconstructFinal}) via (\ref{ADAND1}), (\ref{detA}) gives exact real $2N$-lump solution of of KP-1 equation with integrable boundary (\ref{BoundaryCondition}). We emphasise that  reality and integrable boundary conditions for the constructed solutions $u(x,y,t)$ in general determinant form (\ref{ReconstructFinal}) are exactly satisfied due to the properties (\ref{ADAND1}), (\ref{detA}) of the matrix $A$.

\begin{figure}[h]
\begin{center}
\includegraphics[width=0.50\textwidth, keepaspectratio]{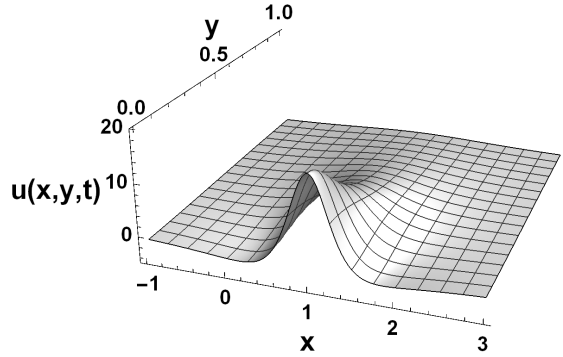}
\parbox[t]{1\textwidth}{\caption{Two-lump solution of KP-1 $u$ (\ref{TwoLumpSolutionKP1})  with parameter $\mu_{10}=1$, $\gamma=1$.}\label{Lump(ImaginarySpectral)KP1}}
\end{center}
\end{figure}

As illustration we calculated two-lump and four-lump exact solutions of KP-1 corresponding to the kernel of the type  $R_0$ (\ref{Kernel3}) for the cases of one  $N=1$ and two $N=2$ paired terms. For one paired  terms in (\ref{Kernel3}) matrix $A$ (\ref{ADAND1}) has the form:
\begin{equation}\label{MatrixA_KP}
\fl A=
\left(
\begin{array}{cc}
         X_1(x,t)-iY_1(y) & -\frac{1}{2\mu_{10}}\\
           \frac{1}{2\mu_{10}} &  X_1(x,t)+iY_1(y) \\
             \end{array}
           \right),\quad
\det A=X^2_1+Y^2_1+\frac{1}{4\mu^2_{10}}
\end{equation}
with
\begin{equation}\label{XY_KP1}
X_1(x,t)=x-12\mu^2_{10}t-\gamma_1,\quad Y_1(y)=2\mu_{10}y.
\end{equation}
Reconstruction formula (\ref{ReconstructFinal}) gives corresponding exact real nonsingular two-lump solution of KP-1 equation with integrable boundary (\ref{BoundaryCondition}):
\begin{equation}\label{TwoLumpSolutionKP1}
u(x,y,t)=2\frac{\partial^2}{\partial x^2}\ln{\det A}=4\frac{Y_1^2-X_1^2+\frac{1}{4\mu^2_{10}}}{\left(X_1^2+Y_1^2+\frac{1}{4\mu^2_{10}}\right)^2}
\end{equation}
which represents localized object moving along $x$-axe with velocity $V_x=12\mu_0^2$.
The graph of this solution on semi-plane $y\geq 0$ is shown on figure (\ref{Lump(ImaginarySpectral)KP1}).

It is interesting to note that to simple kernels $R_{0k}$ corresponding to two separate terms of one pair in $R_0$ (\ref{Kernel3}) with $N=1$, i.e. for kernels
\begin{equation}
\fl R_{01}(\mu,\lambda)=a_1\delta(\mu-i\mu_{10})\delta(\lambda-i\mu_{10}),\quad R_{02}(\mu,\lambda)=a_1\delta(\mu+i\mu_{10})\delta(\lambda+i\mu_{10}),
\end{equation}
due to general formulas (\ref{AD})-(\ref{ReconstructFinal}), correspond the following exact complex-valued one-lump solutions
\begin{equation}\label{ComplexOneLump}
\fl u_1\big|_{\mu_1=i\mu_{10}}=-\frac{2}{\left(X_1(x,t)-iY_1(y)\right)^2},\quad u_2\big|_{\mu_1=-i\mu_{10}}=-\frac{2}{\left(X_1(x,t)+iY_1(y)\right)^2}
\end{equation}
with the same $X_1$ and $Y_1$ (\ref{XY_KP1}) as in (\ref{TwoLumpSolutionKP1}). These solutions (\ref{ComplexOneLump}) have a point singularities at the points $X_1=0$, $Y_1=0$ and did not satisfy to boundary condition (\ref{BoundaryCondition}). The exact two-lump solution (\ref{TwoLumpSolutionKP1}) can be rewritten in the form
\begin{equation}\label{TwoLumpSolutionKP12}
\fl u(x,y,t)=-\frac{2}{\left(X_1(x,t)-i\sqrt{Y^2_1(y)+
\frac{1}{4\mu^2_{10}}}\right)^2}-\frac{2}{\left(X_1(x,t)+
i\sqrt{Y^2_1(y)+\frac{1}{4\mu^2_{10}}}\right)^2}
\end{equation}
of the sum of two simple complex lumps $u_1$ and $u_2$ from (\ref{ComplexOneLump}) with modified (deformed) phase $Y_1\rightarrow Y_{1D}=\sqrt{Y_1^2+\frac{1}{4\mu^2_{10}}}$.
We state that  imposition of boundary condition (\ref{BoundaryCondition}) leads to formation of \,"bound state"\, of two localized simple lumps $u_1$ and $u_2$ given by (\ref{ComplexOneLump})  with modified phase $Y_{1D}=\sqrt{Y_1^2+\frac{1}{4\mu^2_{10}}}$:
\begin{eqnarray}\label{TwoLumpSolutionKP13}
u_{1+2}(x,y,t)=u_1\big|_{\mu_1=i\mu_{10},Y_1\rightarrow Y_{1D}}+u_2\big|_{\mu_1=-i\mu_{10},Y_1\rightarrow Y_{1D}}=\nonumber\\
=-\frac{2}{\left(X_1-iY_{1D}\right)^2}-\frac{2}{\left(X_1+iY_{1D}\right)^2}.
\end{eqnarray}
The "mixture" of two lumps $u_1$ and $u_2$  (\ref{ComplexOneLump}) via the kernel $R_0$ (\ref{Kernel3}) for $N=1$, with one pair of terms with spectral points $\pm i\mu_{10}$, under imposition of boundary condition (\ref{BoundaryCondition}) $u_y|_{y=0}=0$ leads to the formation of eigenmode of the field $u(x,y,t)$ in semi-plane $y\geq0$ in the form of two bounded lumps propagating  on semiplane $y\geq0$  along $x$-axis with velocity $V_x=12\mu^2_{10}$. To this eigenmode corresponds two-lump solution (\ref{TwoLumpSolutionKP1}) or (\ref{TwoLumpSolutionKP12}), (\ref{TwoLumpSolutionKP13}) which is nonlinear superposition of two one-lump solutions (\ref{ComplexOneLump}) very similar to their sums, but with nonlinearly "shifted" phase $Y_1\rightarrow Y_{1D}=\sqrt{Y_1^2+\frac{1}{4\mu^2_{10}}}$.
Zero lines $u=0$ for considered eigenmode of two bounded lumps $u_1$ and $u_2$  (\ref{ComplexOneLump}) are given by equation:
\begin{equation}
X_1^2(x,t)-Y_1^2(y)=\frac{1}{4\mu^2_{10}};
\end{equation}
these resulting from boundary condition (\ref{BoundaryCondition}) zero lines are analogs of Chladni lines for vibrating two-dimensional membranes with fixed boundaries.

\begin{figure}[h]
\begin{center}
\includegraphics[width=0.50\textwidth, keepaspectratio]{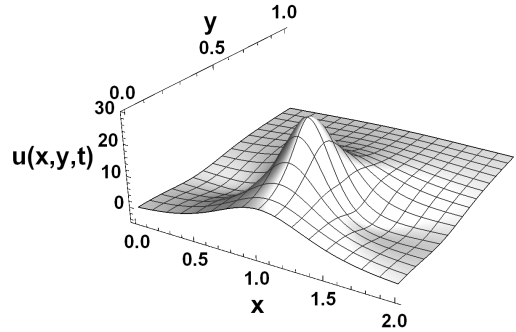}
\parbox[t]{1\textwidth}{\caption{Four-lump solution of KP-1 $u$  with parameters $\mu_{10}=1$, $\mu_{20}=2$, $\gamma_1=\gamma_2=1$.}\label{Lump4(ImaginarySpectral)KP1}}
\end{center}
\end{figure}

In the case of $N=2$ two paired terms in $R_0$ (\ref{Kernel3}) with spectral points $\pm i\mu_{10}$, $\pm i\mu_{20}$ the detailed calculations with general formulae
(\ref{ConditionOnAmplitude})-(\ref{detA}) lead to the following results. Matrix $A$ has the following diagonal $A_D$ and non diagonal $A_F:=A-A_D$ parts:
\begin{eqnarray}\label{ADAND1(1)}
\fl A_{F}=-\frac{i(1-\delta_{kl})}{M_k-M_l}, \quad A_{D}=\textrm{diag}\left(X_1-iY_1,X_1+iY_1;X_2-iY_2,X_2+iY_2\right)\nonumber\\
\fl M_k=(i\mu_{10},-i\mu_{10}, i\mu_{20},-i\mu_{20}),\quad X_k(x,t)=x-12\mu^2_{k0}t-\gamma_k,\, Y_k(y)=2\mu_{k0}y.
\end{eqnarray}
here $k,l=1,\ldots,4$ and $X_k(x,t)=\overline X_k$, $Y_k(y)=\overline Y_k$.
For $\det A$ we described the following expression:
\begin{eqnarray}\label{detA_KP}
\det{A}=\left|\Phi_1(\mu_{10})\Phi_2(\mu_{20})+\frac{1}{\left(\mu_{10}-
\mu_{20}\right)^2}\right|^2+
\frac{\left|\mu_{10}\Phi_1(\mu_{10})+
\mu_{20}\Phi_2(\mu_{20})\right|^2}{\mu_{10}\mu_{20}\left(\mu_{10}+\mu_{20}\right)^2}+ \nonumber \\
+\frac{1}{4\mu^2_{10}}\left(\frac{\mu_{10}-\mu_{20}}{\mu_{10}+\mu_{20}}\right)^2\left|\Phi_1(\mu_{20})\right|^2
+\frac{1}{4\mu^2_{20}}\left(\frac{\mu_{10}-\mu_{20}}{\mu_{10}+
\mu_{20}}\right)^2\left|\Phi_1(\mu_{10})\right|^2+\nonumber \\
+\frac{1}{\left(\mu_{10}+\mu_{20}\right)^4}+\frac{1}{16\mu^2_{10}\mu^2_{20}},
\end{eqnarray}
here
\begin{equation}\label{XX}
\Phi_1(\mu_{10}):=X_1(x,t)-iY_1(y),\quad \Phi_2(\mu_{20}):=X_2(x,t)-iY_2(y)
\end{equation}
and $X_k(x,t)$, $Y_k(y)$, $k=1,2$ are given by (\ref{X_kp1}).
Reconstruction formula (\ref{ReconstructFinal}) gives the corresponding real four-lump solution with integrable boundary condition $u_y\big|_{y=0}=0$ which is nonsingular if $\mu_{10}\mu_{20}>0$. Under the limits $\Phi_1(\mu_{10})=\textrm{const}$, $\Phi_2(\mu_{20})\rightarrow \infty$ and $\Phi_1(\mu_{10})\rightarrow \infty$, $\Phi_2(\mu_{20})=\textrm{const}$ we derived from (\ref{detA_KP}):
\begin{eqnarray}\label{detALimits}
\det A\big|_{\Phi_1(\mu_{10})=\textrm{const},\,\Phi_2(\mu_{20})\rightarrow \infty}\rightarrow\left|\Phi_2(\mu_{20})\right|^2\left(\left|\Phi_1(\mu_{10})\right|^2+\frac{1}{4\mu^2_{10}}\right),\\
\det A\big|_{\Phi_1(\mu_{10})\rightarrow \infty,\,\Phi_2(\mu_{20})=\textrm{const}}\rightarrow\left|\Phi_1(\mu_{10})\right|^2
\left(\left|\Phi_2(\mu_{20})\right|^2+\frac{1}{4\mu^2_{20}}\right),
\end{eqnarray}
and via (\ref{ReconstructFinal}) the corresponding two-lump solutions of the type (\ref{TwoLumpSolutionKP1}). Here in (\ref{detALimits}) $\left|\Phi_1(\mu_{10})\right|^2+\frac{1}{\mu^2_{10}}$ and $\left|\Phi_2(\mu_{20})\right|^2+\frac{1}{\mu^2_{20}}$ are exactly the determinants of the type (\ref{MatrixA_KP}) for  two-lump solutions of the type (\ref{TwoLumpSolutionKP1}) with corresponding pairs of spectral points $(i\mu_{10},-i\mu_{10})$ and $(i\mu_{20},-i\mu_{20})$.
So four-lump solution given by reconstruction formula (\ref{AD})-(\ref{ReconstructFinal}), with matrix $A$ and $\det A$ in (\ref{ADAND1(1)})-(\ref{XX}), under the condition $\mu_{10}\mu_{20}>0$, represents nonsingular eigenmode of the field $u(x,y,t)$ in semi-plane $y\geq0$ in the form of four simple bounded lumps of the type (\ref{ComplexOneLump}).
The graph of this solution on semi-plane $y\geq 0$ is shown on figure (\ref{Lump4(ImaginarySpectral)KP1}).

\section{Multi-lump solutions of KP-2 equation with integrable boundary condition}
\label{Section_4}
First, we considered the kernel $R_0$ with pure imaginary spectral points $\pm i\mu_{k0}$. As in Section 3 for KP-1 to restriction (\ref{BoundaryConditionWeakField}) from boundary condition (\ref{BoundaryCondition})
satisfies delta-form kernel $R_0$ (\ref{kernel1}) with paired terms
\begin{equation}\label{Kernel4}
\fl R_0(\mu,\overline\mu;\lambda,\overline\lambda)=\sum\limits_{k=1}^N\left(a_{1k}\delta(\mu-i\mu_{k0})\delta(\lambda-i\mu_{k0})+
a_{2k}\delta(\mu+i\mu_{k0})\delta(\lambda+i\mu_{k0})\right)
\end{equation}
and with equal amplitudes $a_{1k}$ and $a_{2k}$:
\begin{equation}
a_{1k}=a_{2k}=a_k.
\end{equation}
We also further demonstrated that reality condition $\overline u=u$ is satisfied for real amplitudes $a_k$:
\begin{equation}
a_{1k}=a_{2k}=a_k=\overline a_k.
\end{equation}
We rewrote the kernel $R_0$ (\ref{Kernel4}) in form (\ref{KernelGenForm}), for convenience of application of general determinant formula (\ref{ReconstructFinal}), using the sets of amplitudes $A_k$ and spectral points $M_k$ $(k=1,2,\ldots,2N)$:
\begin{eqnarray}\label{SetsOfParametr}
\fl\left(A_1,A_2,\ldots,A_{2N-1},A_{2N}\right)=(a_1,a_1;a_2,a_2;\ldots;a_N,a_N), \nonumber
\\
\fl\left(M_1,M_2,\ldots,M_{2N-1},M_{2N}\right)=
(i\mu_{10},-i\mu_{10};i\mu_{20},-i\mu_{20};\ldots;i\mu_{N0},-i\mu_{N0}).
\end{eqnarray}
Using (\ref{SetsOfParametr}) we derived for the matrix $A$ given by (\ref{XtildeX}) the following expression:
\begin{equation}\label{MatrixA}
A_{kl}= \Phi_k(M_k)\delta_{kl}-\frac{i(1-\delta_{kl})}{M_k-M_l}, \quad (k,l=1,\ldots,2N)
\end{equation}
with phases
\begin{equation}
\Phi_k(M_k=\pm i\mu_{k0}):=x-12\mu_{k0}^2t-\frac{\pi}{2a_k}\pm2\mu_{k0}y:=X_k(x,t)\pm Y_k(y),
\end{equation}
here
\begin{equation}\label{XY}
\fl X_k(x,t):=x-12\mu_{k0}^2t-\gamma_k=\overline{X}_k(x,t), \quad  Y_k(y):=2\mu_{k0}y=\overline{Y}_k(y);\quad \gamma_k:=\frac{\pi}{2a_k}=\overline{\gamma_k}.
\end{equation}
Due to (\ref{MatrixA})-(\ref{XY}), under requirement of real amplitudes $a_k=\overline {a_k}$ in (\ref{Kernel4}), all  elements of matrix $A$ are real and  therefore
determinant of matrix $A$ is real:
\begin{equation}
\overline{\det A}=\det A.
\end{equation}
Moreover the matrix $A$ has the following diagonal $A_D$ and non diagonal $A_F:=A-A_D$ parts:
\begin{eqnarray}\label{ADAND(1)}
A=A_D+A_F,\quad A_F=-\frac{i(1-\delta_{kl})}{M_k-M_l},\quad (k,l=1,\ldots,2N), \nonumber
\\
A_{D}=\textrm{diag}\left(X_1+Y_1,X_1-Y_1;\ldots;X_N+Y_N, X_N-Y_N\right).
\end{eqnarray}
Due to (\ref{ADAND(1)}) $\det A$ is even function of $y$, i.e.
\begin{equation}\label{ReconstructFinal2}
\det A=\det A(y^2),
\end{equation}
so the boundary condition $u_y\big|_{y=0}=0$ due to (\ref{ReconstructFinal2}) is also satisfied, and consequently general determinant formula (\ref{ReconstructFinal}) gives corresponding to the kernel (\ref{Kernel4}) new class of exact real $2N$-lump solutions of KP-2 equation with integrable boundary condition (\ref{BoundaryCondition}).

\begin{figure}[h]
\begin{center}
\includegraphics[width=0.50\textwidth, keepaspectratio]{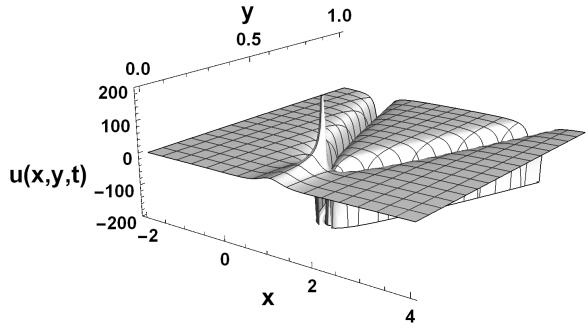}
\parbox[t]{1\textwidth}{\caption{Two-lump solution of KP-2 $u$ (\ref{KP2TwoLump})  with parameter $\mu_{10}=1$, $\gamma=1$.}\label{pictLump(ImaginarySpectral)KP2}}
\end{center}
\end{figure}

For the simplest two-lump solution corresponding to one pair $(N=1)$ of spectral points $(i\mu_{10},-i\mu_{10})$ in kernel $R_0$ (\ref{Kernel4}) matrix $A$ has the form:
\begin{eqnarray}
A= \left(
  \begin{array}{cc}
    X_1(x,t)+Y_1(y) & -\frac{1}{2\mu_{10}} \\
    \frac{1}{2\mu_{10}} & X_1(x,t)-Y_1(y) \\
  \end{array}
\right),\nonumber
\\
X_1(x,t)=x-12\mu^2_{10}t-\gamma_1,\quad Y_1(y)=2\mu_{10}y.
\end{eqnarray}
For $\det A$ we obtained:
\begin{equation}
\det A=X_1^2-Y_1^2+\frac{1}{4\mu^2_{10}}.
\end{equation}
The reconstruction formula (\ref{ReconstructFinal}) gives real singular two-lump solution:
\begin{equation}\label{KP2TwoLump}
u(x,y,t)=2\frac{\partial^2}{\partial x^2}\ln\det A=-4\frac{X_1^2(x,t)+Y^2_1(y)-\frac{1}{4\mu^2_{10}}}{\left(X_1^2(x,t)-
Y^2_1(y)+\frac{1}{4\mu^2_{10}}\right)^2},
\end{equation}
with singularity line in the form of hyperbolas
\begin{equation}\label{SingularityLine}
Y^2_1(y)-X_1^2(x,t)=\frac{1}{4\mu^2_{10}},
\end{equation}
and zero line - in the form of the circle
\begin{equation}\label{ZeroLine}
Y^2_1(y)+X_1^2(x,t)=\frac{1}{4\mu^2_{10}}
\end{equation}
on the plane $(X_1,Y_1)$.
Both lines of singularities (\ref{SingularityLine}) and zero line (\ref{ZeroLine}) are propagate along the $x$-axe with velocity $V_x=12\mu_{10}^2$.
The graph of this solution (\ref{KP2TwoLump}) on semi-plane $y\geq 0$ is shown on figure (\ref{pictLump(ImaginarySpectral)KP2}).

It is interesting to note that for simple kernels $R_{01}$ and $R_{02}$ corresponding to two separate terms of one pair in $R_0$ (\ref{Kernel4}) with $N=1$, i.e. for kernels with one delta-form term (one product of two delta functions),
\begin{equation}\label{twoKernels}
\fl R_{01}(\mu,\lambda)=a_1\delta(\mu-i\mu_{10})\delta(\lambda-i\mu_{10}),\quad R_{02}(\mu,\lambda)=a_1\delta(\mu+i\mu_{10})\delta(\lambda+i\mu_{10})
\end{equation}
by the use of general formulas (\ref{AD})-(\ref{ReconstructFinal}) we derived exact singular one-lump solutions:
\begin{equation}\label{OneLumpKP2}
u_1\big|_{\mu=i\mu_{10}}=-\frac{2}{\left(X_1(x,t)-Y_1(y)\right)^2},\quad u_2\big|_{\mu=-i\mu_{10}}=-\frac{2}{\left(X_1(x,t)+Y_1(y)\right)^2}.
\end{equation}
Two-lump solution (\ref{KP2TwoLump}) can be represented in the form:
\begin{equation}\label{TwoOneLumpKP2}
\fl u(x,y,t)=-\frac{2}{\left(X_1(x,t)-\sqrt{Y^2_1(y)-
\frac{1}{4\mu^2_{10}}}\right)^2}-\frac{2}{\left(X_1(x,t)+\sqrt{Y^2_1(y)-
\frac{1}{4\mu^2_{10}}}\right)^2}
\end{equation}
very similar to the sum of $u_1$ and $u_2$ from (\ref{OneLumpKP2}) but with modified (or deformed) phase
\begin{equation}\label{DeformedPhase}
Y_{1D}=\sqrt{Y_1^2-\frac{1}{4\mu^2_{10}}},
\end{equation}
 i.e.
\begin{equation}\label{property}
u(x,y,t)=\left(u_1+u_2\right)\big|_{Y_1(y)\rightarrow Y_{1D}:=\sqrt{Y^2_1-\frac{1}{4\mu^2_{10}}}}.
\end{equation}
The kernel $R_0=R_{01}+R_{02}$ as the sum of kernels (\ref{twoKernels}) leads to the exact two-lump solution (\ref{KP2TwoLump}) or (\ref{TwoOneLumpKP2})  with property (\ref{property}): exact two-lump solution (\ref{TwoOneLumpKP2}) is  nonlinear superposition of $u_1$ and $u_2$ with modified phase $Y_1\rightarrow Y_{1D}$.
The fulfillment of boundary condition (\ref{BoundaryCondition}) leads to formation of \,"eigen-mode"\, of field  $u(x,y,t)$  at semi-plane $y\geq 0$ represented by two-lump solution $u(x,y,t)$ (\ref{KP2TwoLump})  in the form of two (\ref{property}) bounded with each other one-lump solutions (\ref{OneLumpKP2}).

The kernel
\begin{equation}\label{Kernel5}
\fl R_0(\mu,\overline\mu;\lambda,\overline\lambda)=\sum\limits_{k=1}^N\left(a_{1k}\delta(\mu-\mu_{k})\delta(\lambda-\mu_{k})+
a_{2k}\delta(\mu+\mu_{k})\delta(\lambda+\mu_{k})\right)
\end{equation}
with paired terms with real $\overline{\mu_k}=\mu_k$ spectral points corresponds to another interesting class of exact multi-lump solutions of KP-2 equation.
The restriction (\ref{BoundaryConditionWeakField}) from boundary condition (\ref{BoundaryCondition}) leads to the equality of amplitudes $a_{1k}=a_{2k}=a_k$,  reality condition $\overline u= u$ is satisfied in this considered case by real amplitudes $a_{k}$:
\begin{equation}
a_{1k}=a_{2k}=a_k=\overline a_k.
\end{equation}
It is convenient to rewrite kernel $R_0$ (\ref{Kernel5}) in the form (\ref{KernelGenForm}) to apply general determinant formula (\ref{ReconstructFinal}). Here, the sets of amplitudes $A_k$ and spectral points $M_k$ $(k=1,2,\ldots,2N)$ are given by:
 \begin{eqnarray}\label{setA&M_KP2}
 \left(A_1,A_2,\ldots,A_{2N-1},A_{2N}\right)=\left(a_1,a_1;a_2,a_2,\ldots,a_N, a_N\right),\nonumber \\
  \left(M_1,M_2,\ldots,M_{2N-1},M_{2N}\right)=
  \left(\mu_1,-\mu_1;\mu_2,-\mu_2,\ldots,\mu_N,-\mu_N\right).
 \end{eqnarray}
The matrix $A$ due to (\ref{AD}) and (\ref{setA&M_KP2}) has the following form:
 \begin{equation}\label{ADAND(KP2)}
 A=\Phi_k(M_k)\delta_{kl}-\frac{i(1-\delta_{kl})}{M_k-M_l}, \quad(k,l=1,\ldots,2N)
 \end{equation}
 with diagonal part
 \begin{equation}
 A_D=\textrm{diag}(X_1-iY_1, X_1+iY_1;\ldots; X_N-iY_N,X_N+iY_N).
 \end{equation}
 Here
 \begin{eqnarray}\label{XY_KP2}
 \Phi_k(\pm \mu_k)=x+12\mu^2_k t-\gamma_k\mp 2i\mu_k y:=X_k(x,t)\mp iY_k(y),\nonumber \\ \gamma_k:=\frac{\pi}{2a_k}=\overline{\gamma_k};\quad
 \overline{X_k(x,t)}=X_k(x,t),\quad \overline{Y_k(y)}=Y_k(y).
  \end{eqnarray}
  Once again due to (\ref{ADAND(KP2)})-(\ref{XY_KP2}) $\det A$ is real and even function of $y$:
  \begin{equation}
  \overline{\det A}=\det A=\det A(y^2),
  \end{equation}
  and general determinant formula (\ref{ReconstructFinal}) gives new class of exact real but singular 2N-lump solutions of KP-2 with integrable boundary condition $u_y\big|_{y=0}=0$.

\begin{figure}[h]
\begin{center}
\includegraphics[width=0.50\textwidth, keepaspectratio]{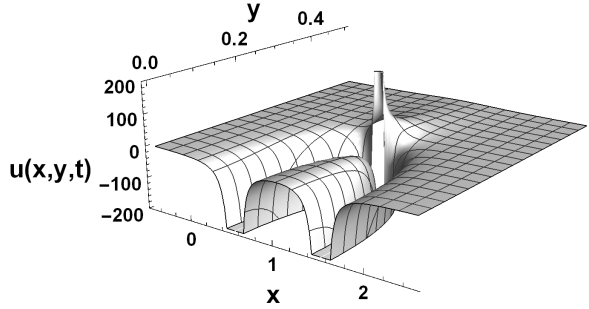}
\parbox[t]{1\textwidth}{\caption{Two-lump solution of KP-2 $u$ (\ref{TwoLumpKP2(1)}) with parameter $\mu_{1}=1$, $\gamma=1$.}\label{pictKP2TwoLump}}
\end{center}
\end{figure}

 The simplest two-lump solution corresponds to one pair of real spectral points $(\mu_1,-\mu_1)$ in kernel $R_0$ (\ref{Kernel5}). Matrix $A$ and $\det A$ in present case:
  \begin{equation}
A=  \left(
    \begin{array}{cc}
      X_1-iY_1 & -\frac{i}{2\mu_1} \\
      \frac{i}{2\mu_1} & X_1+iY_1 \\
    \end{array}
  \right),
\end{equation}
  \begin{equation}
\fl  X_1(x,t)=x+12\mu_1^2t-\gamma_1,\quad Y_1(y):=2\mu_1 y; \quad
  \det A=X_1^2(x,t)+Y_1^2(y)-\frac{1}{4\mu_1^2}.
\end{equation}
The exact solution in this case is given by the formula:
 \begin{equation}\label{TwoLumpKP2(1)}
 u(x,y,t)=2\frac{\partial^2}{\partial x^2}\ln\det A=-4\frac{X_1^2(x,t)-Y_1^2(y)+\frac{1}{4\mu_1^2}}{\left(X_1^2+Y_1^2-\frac{1}{4\mu^2_1}\right)^2}
 \end{equation}
 with line of singularities in the form of circle
 \begin{equation}
 X_1^2(x,t)+Y_1^2(y)=\frac{1}{4\mu^2_1},
 \end{equation}
 and with zero lines - in the form of hyperbolas
 \begin{equation}
 Y_1^2(y)-X_1^2(x,t)=\frac{1}{4\mu^2_1}
 \end{equation}
 on the plane $(X_1,Y_1)$.
The graph of this real singular solution in semi-plane $y\geq 0$ is shown on figure (\ref{pictKP2TwoLump}).

It is interesting to note that to simple kernels $R_{01}$ and $R_{02}$ with one delta-term:
\begin{equation}\label{twoKernels(2)}
R_{01}(\mu,\lambda)=a_1\delta(\mu-\mu_{1})\delta(\lambda-\mu_{1}),\quad R_{02}(\mu,\lambda)=a_1\delta(\mu+\mu_{1})\delta(\lambda+\mu_{1})
\end{equation}
general formulas (\ref{AD})-(\ref{ReconstructFinal}) lead to exact complex one-lump solutions $u_1$, $u_2$:
\begin{equation}\label{TwoOneLumpKP2(1)}
\fl u_1\big|_{\mu=\mu_{1}}=-\frac{2}{\left(X_1(x,t)-iY_1(y)\right)^2},\quad u_2\big|_{\mu=-\mu_{1}}=-\frac{2}{\left(X_1(x,t)+iY_1(y)\right)^2}.
\end{equation}
The exact solution (\ref{TwoLumpKP2(1)}) can be represented as the sum of $u_1$ and $u_2$ (\ref{TwoOneLumpKP2(1)}) but with modified phase $Y_1(y)\rightarrow Y_{1D}=\sqrt{Y_1^2-\frac{1}{4\mu^2_{1}}}$:
\begin{eqnarray}\label{OneLumpKP2(1)}
\fl u(x,y,t)=-\frac{2}{\left(X_1(x,t)-i\sqrt{Y^2_1(y)-\frac{1}{4\mu_1^2}}\right)^2}-
\frac{2}{\left(X_1(x,t)+i\sqrt{Y^2_1(y)-\frac{1}{4\mu_1^2}}\right)^2}=\nonumber\\
\fl=(u_1+u_2)\big|_{Y_1(y)\rightarrow Y_{1D}(y)=\sqrt{Y_1^2-\frac{1}{4\mu^2_{1}}}}.
\end{eqnarray}
Once again, as for all considered in present paper exact two-lump solutions of KP-1 and KP-2 equations, the fulfilment of boundary condition (\ref{BoundaryCondition}) via nonlinear superposition of two bounded with each other simple lumps (\ref{TwoOneLumpKP2(1)}) leads to the formation of eigenmode (\ref{TwoLumpKP2(1)}) or (\ref{OneLumpKP2(1)}) of field $u(x,y,t)$ in semi-plane $y\geq0$.

\section{Conclusions}
\label{Section_7}
\setcounter{equation}{0}

In the present paper we described new classes of exact real multi-lump solutions of KP-1,2 equations with integrable boundary condition $u_{y}\big|_{y=0}=0$ and developed general scheme for this in framework of $\overline\partial$-dressing method. We demonstrated that reality $u=\overline u$ and boundary conditions for solutions $u$ can be effectively satisfied exactly by the use of general determinant formula,
this leads to some restrictions on parameters of solutions, i.e. on amplitudes $A_k$ and spectral points $\mu_k$, $\lambda_k$ of delta-form kernel $R_0$ of $\overline\partial$-problem (\ref{di_problem1}).

We presented the simplest examples of real two-lump exact solutions as illustrations. We calculated as examples of nonsingular and also singular solutions which  have point and line singularities and Chladny-type zero lines $u(x,y,t)=0$, such solutions  belong to the class of solutions with integrable boundary condition.

We demonstrated the effectiveness of $\overline\partial$-dressing method in calculation of multi-lump solutions with integrable boundary conditions. The imposition on the field $u$ of boundary condition (\ref{BoundaryCondition}) leads to formation of bounded with each other simple lumps or the eigenmodes of the field $u(x,y,t)$  on semi-plane $y\geq 0$. Such eigenmodes of "coherently connected" with each other simple lumps  propagate with some velocity along $x$-axis.

The developed in present paper procedure for calculation via $\overline\partial$-dressing of new classes of exact real multi-lump solutions can be effectively applied to all other integrable (2+1)-dimensional nonlinear equations, some of these studies are currently in progress and the results will be published elsewhere.

Possible physical applications of calculated in the present paper exact multi-lump solutions of KP equation should be noted.  KP equation can be applied for description of fluids flows in thin films on inclined surfaces in Earth gravity field. There may be, due to some specific experimental boundary conditions, some kind of fluid \,"excitations"\, in such films, in the form of "coherently connected" with each other simple lumps, that could be observed by hydrodynamics experimentalists.

\section*{References}
\setcounter{equation}{0}


\begin{thebibliography}{99}

\bibitem{KadPetv} B. B. Kadomtsev, V. I. Petviashvili, "On the stability of solitary waves in weakly dispersing media", Dokl. Akad. Nauk SSSR, 192:4 (1970), 753–756.

\bibitem{Dryuma} V. S. Dryuma, Analitic solution of the two-dimemsional Korteveg–de Vries (KdV) equation, Sov. Phys. JETF Lett., 19 (1974), 387–388.

\bibitem{AblowitzClarkson} M.J.Ablowitz, P.A. Clarkson. Solitons, Nonlinear Evolution Equations and Inverse Scattering. London Mathematical Society Lecture Note Series. Cambridge University Press, 1991.

\bibitem{AblowitzSegur} M.J.Ablowitz, H.Segur. Solitons and the Inverse Scattering Transform. Series: SIAM Studies in Applied Mathematics. Society for Industrial and Applied Mathematics 1981.

\bibitem{NovikovManakov} S.P. Novikov, S.V. Manakov, L.P. Pitaevskii, V.E. Zakharov. Theory of Solitons: The Inverse Scattering Method. Series: Monographs in Contemporary Mathematics. Springer US 1984.

\bibitem{ZakharovShabat} V.E.Zakharov, A.B.Shabat. A scheme for integrating the nonlinear equations of mathematical physics by the method of the inverse scattering problem I.  Funct Anal Its Appl (1974) 8: 226. https://doi.org/10.1007/BF01075696

\bibitem{ZakharovShabat2} V.E.Zakharov, A.B.Shabat. Integration of nonlinear equations of mathematical physics by the method of inverse scattering II.  Funct Anal Its Appl (1979) 13: 166. https://doi.org/10.1007/BF01077483

\bibitem{Manakov} S.V. Manakov. The inverse scattering transform for the time-dependent Schrodinger equation and Kadomtsev-Petviashvili equation // Physica D. 1981. Vol. 3 (1-2),  pp. 420-427. doi: 10.1016/0167-2789(81)90145-7

\bibitem{Fokas&Ablowitz}	A.S. Fokas, M.J. Ablowitz, The inverse scattering transform for multidimensional (2+1) problems // Lecture Notes in Physics, vol. 189, p.137-183 Nonlinear Phenomena. [Proceedings of the CIFMO School and Workshop held at Oaxtepec, Mexico November 29 - December 17, 1982.]

\bibitem{Zakharov&Manakov} V. E. Zakharov, S. V. Manakov, "Construction of higher-dimensional nonlinear integrable systems and of their solutions", Funktsional. Anal. i Prilozhen., 19:2 (1985), 11–25; Funct. Anal. Appl., 19:2 (1985), 89–101. https://doi.org/10.1007/BF01078388

\bibitem{Beals&Coifman1} R. Beals, R.R. Coifman. The D-bar approach to inverse scattering and nonlinear evolutions// Physica D. 1986. Vol. 18 (1-3), pp. 242-249. doi: 10.1016/0167-2789(86)90184-3

\bibitem{Beals&Coifman2} R. Beals, R.R. Coifman.  Linear spectral problems, non-linear equations and the $\overline\partial$-method. Inverse Problems. 1989. Vol. 5 (87), pp. 87-130. doi: 10.1088/0266-5611/5/2/002

\bibitem{Sklyanin} E. K. Sklyanin, Boundary conditions for integrable equations,  Funktsional. Anal. i Prilozhen., 21:2 (1987),  86–87; Funct. Anal. Appl., 21:2 (1987), 164–166. https://doi.org/10.1007/BF01078038

\bibitem{Habibullin}  E. V. Gudkova, I. T. Habibullin, Kadomtsev–Petviashvili Equation on the Half-Plane, Theoret. and Math. Phys., 140:2 (2004), 1086–1094, https://doi.org/10.1023/B:TAMP.0000036539.35565.1f

\bibitem{Habibullin2}	Habibullin I. T., Gudkova E. V., Boundary Conditions for Multidimensional Integrable Equations, Funktsional. Anal. i Prilozhen., 38:2 (2004), 71–83; Funct. Anal. Appl., 38:2 (2004), 138–148. https://doi.org/10.4213/faa109

\bibitem{HabibullinSinGSchr} I.T. Khabibullin, Boundary conditions for nonlinear equations compatible with integrability, Theor Math Phys (1993) 96: 845. https://doi.org/10.1007/BF01074113

\bibitem{HabibullinSineG}  I.T. Khabibullin, Sine-Gordon equation on the semi-axis. Theor Math Phys (1998) 114: 90. https://doi.org/10.1007/BF02557111

\bibitem{HabibullinAdlerShabat} V.E. Adler, L.T. Habibullin, A.B. Shabat, Boundary value problem for the KDV equation on a half-line. Theor Math Phys (1997) 110: 78.
https://doi.org/10.1007/BF02630371

\bibitem{HabibullinKDV} I.T. Khabibullin, KDV equation on a half-line with the zero boundary condition. Theor Math Phys (1999) 119: 712. https://doi.org/10.1007/BF02557381

\bibitem{Vereshchagin} V. L. Vereshchagin, Integrable boundary conditions for (2+1)-dimensional models of mathematical physics. Theor Math Phys (2012) 171: 792.  https://doi.org/10.1007/s11232-012-0075-9

\bibitem{FokasBook} A.S. Fokas. A Unified Approach to Boundary Value Problems. Society for Industrial and Applied Mathematics. 2008. doi:10.1137/1.9780898717068.

\bibitem{Zakharov} V.E. Zakharov.  Commutating operators and nonlocal  problem // Plasma theory and Nonlinear and turbulent processes in Physics / Ed. by Erokhin N.S., Zakharov V.E., Sitenko A.G., Chernousenko V.M. Bar'yakhtar V.G. Kiev: Naukova Dumka, 1988. Vol. 1, pp. 152.

\bibitem{Bogdanov&Manakov}	L.V. Bogdanov, S.V.Manakov. The non-local  problem and (2+1)-dimensional soliton equations // Journal of Physics A. 1988. Vol. 21 (10), pp. 537-544. doi:10.1088/0305-4470/21/10/001

\bibitem{KonopelchenkoInvProblems}	B.G. Konopelchenko.  The two-dimensional second-order differential spectral problem: compatibility conditions, general BTs and integrable equations // Inverse Problems  1988  4 p.151. https://doi.org/10.1088/0266-5611/4/1/013

\bibitem{Zakharov90} V.E. Zakharov On the dressing method // Inverse Methods in Action / Ed. By P.C.Sabatier. Springer, 1990, pp. 602.

\bibitem{KonopelchenkoBook1} B.G. Konopelchenko. Introduction to Multidimensional Integrable Equations: The Inverse Spectral Transform in 2+1 Dimensions. New York: Plenum Press, 1992.

\bibitem{KonopelchenkoBook2} B.G. Konopelchenko. Solitons in Multidimensions: Inverse Spectral Transform Method. Singapore: World Scientific, 1993.








\end{thebibliography}
\end{document}